# Filtering material properties to improve FFT-based methods for numerical homogenization

Lionel Gélébart, Franck Ouaki

CEA Saclay, DEN/DMN/SRMA, 91191, GIF/YVETTE, FRANCE

Email : lionel.gelebart@cea.fr

Abstract:

FFT-based solvers introduced in the 1990's for the numerical homogenization of heterogeneous elastic materials have been extended to a wide range of physical properties. In parallel, alternative algorithms and modified discrete Green operators have been proposed to accelerate the method and/or improve the description of the local fields. In this short note, filtering material properties is proposed as a third complementary way to improve FFT-based methods. It is evidenced from numerical experiments that, the grid refinement and consequently the computation time and/or the spurious oscillations observed on local fields can be significantly reduced. In addition, while the Voigt and Reuss filters can improve or deteriorate the method depending on the microstructure, a stiff inclusion within a compliant matrix or the reverse, the proposed '2-layers' filter is efficient in both situations. The study is proposed in the context of linear elasticity but similar results are expected in a different physical context (thermal, electrical…).

Keywords :

Numerical homogenization; FFT; Spectral method; Filter; Multilayer model; Spurious oscillations

1 – Introduction

The spectral methods based on the Lippmann-Schwinger equation associated to a discrete Green operator are efficient numerical iterative methods devoted to the evaluation of physical properties of heterogeneous unit cells submitted to periodic boundary conditions. If the method proposed initially [9] suffers from various drawbacks, different propositions have been made to improve it. These propositions can be classified into two categories. The first category [3, 5, 12, 6, 4] operates on the iterative algorithm to reduce the number of iterations until convergence, without changing the numerical solution, while the second category [1, 11] modifies the discrete Green operator to improve both the numerical solution fields and the convergence properties. Actually, spurious oscillations are commonly observed on the numerical solution fields when using the discrete Green operator proposed initially.

The purpose of this short note is to propose the filtering of material properties as a complementary way to improve these methods without any modification of the iterative algorithm or of the discrete Green operator. The idea is inspired by a previous work [10, 2, 1] assigning homogenized properties to heterogeneous finite elements (*i.e.* crossed by an interface between two materials), which can be regarded as a mechanical filter which size is the same as the element size. Here the effect of the filter radius is considered and a new mechanical filter based on a multilayer homogenization rule is proposed in addition to the classical Voigt and Reuss rules. The benefits of this approach, evidenced by a simple numerical experiment, are the improvement of the spatial convergence properties as well as the reduction of spurious oscillations. In this example, for a given convergence criterion, the required spatial resolution can be divided by a factor between 2.6 and 7.2, corresponding to a reduction of the problem size by a factor between 6.7 and 51.8 in 2D (and between 17.5 and 373 if extrapolated to 3D). Thus, reducing the problem size, filtering materials properties decreases the computation time. Moreover, this does not strongly affect the convergence properties of the iterative algorithm, here the classical fixed-point algorithm [8].

2 – Filtering mechanical properties

The exact problem to solve is the evaluation of the strain (stress and displacement) field(s) within a heterogeneous unit cell $\Omega$ described by its stiffness tensor field $c$ and submitted to a macroscopic strain $E$ with periodic boundary conditions. The exact solution is given by the following equations from which a fixed-point algorithm can be derived [8]:

$$\varepsilon(x) = -(\Gamma_0 * \tau)(x) + E \qquad \text{for } x \in \Omega \qquad (1\text{-a})$$
$$\tau(x) = (c(x) - c_0) : \varepsilon(x) \qquad \text{for } x \in \Omega \qquad (1\text{-b})$$

$\Gamma_0$ is the periodic Green operator for a homogeneous medium of stiffness $c_0$. The convolution product $*$ can be written in Fourier space as follows:

$$\begin{cases} \hat{\varepsilon}(k_a) = -\hat{\Gamma}_0(k_a) : \hat{\tau}(k_a) & \text{for } a \in \mathbb{Z}^3 \text{ and } a \neq 0 \\ \hat{\varepsilon}(0) = E \end{cases} \qquad (2)$$

Usually, in order to obtain an approximate solution to the problem on a grid of size $N_1 \times, N_2 \times, N_3$, the polarization $\tau$ (equation 1-b) is simply evaluated at grid points $x_\alpha$, then a Fast Fourier Transform (FFT) is used to evaluate $\hat{\tau}(k_a)$ for $a \in J \ (= \prod_{d=1,3}\{-(N_d - 1)/2, (N_d - 1)/2\}$, for odd discretizations), then the Green operator is applied on a truncated Fourier space (equation 2 with $a \in J$ ) and finally the strain at grid points $x_\alpha$ is deduced from an inverse FFT. A new polarization field can be evaluated from this strain field and so on.

However, using $c(x_\alpha)$ in equation 1-b is quite arbitrary. Instead, we propose to work with a locally filtered homogenized behavior $\tilde{c}$ defined by a uniform filter of shape $\omega$ centered around $x_\alpha$, acting on phase indicator functions to define phase volume fractions, and a homogenization rule (HR) accounting for these volume fractions and the corresponding behaviors : $\tilde{c}(x_\alpha) = HR(c, \omega(x_\alpha))$.

When considering sufficiently refined grids, most of the filters $\omega(x_\alpha)$ consist of a single phase and $\tilde{c}(x_\alpha) = c(x_\alpha)$. When at least two phases belong to $\omega(x_\alpha)$, their volume fractions $f_i(x_\alpha)$ are evaluated (for example by averaging indicator functions evaluated on a refined grid) and Voigt or Reuss homogenization rules can be used as proposed previously in a slightly different context [2] :

$$\tilde{c}(x_\alpha) = Voigt(c, \omega(x_\alpha)) = \sum_{i=1}^{Nphase} f_i(x_\alpha) c_i \qquad (3)$$

$$\tilde{c}(x_\alpha) = Reuss(c, \omega(x_\alpha)) = \left(\sum_{i=1}^{Nphase} f_i(x_\alpha) c_i^{-1}\right)^{-1} \qquad (4)$$

In addition, a new homogenization rule based on the solution of a two-phase multilayered microstructure is proposed when two phases, separated by an interface, belong to $\omega(x_\alpha)$. In that case, in addition to the volume fractions $f_1(x_\alpha)$ and $f_2(x_\alpha)$, a planar approximation of the interface crossing the filter must be evaluated. Using P and OP respectively for the "in-plane" and "out-of-plane" components (with respect to the approximate planar interface) of the stress σ and strain ε tensors, the interface and averaging conditions read :

$$\begin{cases} \sigma_1^{OP} = \sigma_2^{OP} = \sigma^{OP} \\ \varepsilon_1^P = \varepsilon_2^P = \varepsilon^P \end{cases} \qquad (5\text{-a})$$

$$\begin{cases} f_1 \sigma_1^P + f_1 \sigma_2^P = \sigma^P \\ f_2 \varepsilon_1^{OP} + f_2 \varepsilon_2^{OP} = \varepsilon^{OP} \end{cases} \qquad (5\text{-b})$$

Solving this set of equations leads to the definition of the "2-layer" homogenized stiffness tensor.

In the following, the three homogenization rules are applied with different filter radii, to a simple 2D microstructure, a disk (radius $\frac{\pi}{15} \sim 0.21$) centered within a square unit cell (size 1x1). Both inclusion and matrix have an elastic isotropic behavior with the same Poisson coefficient, $\nu = 0.3$, and the elastic contrast is given by the Young modulus ratio, $E_{inclusion}/E_{matrix}$. The loading is a uniaxial average strain (1%). The classical fixed-point algorithm described in section 1 is used with a $10^{-4}$ stress equilibrium criterion (see definition in [9]). The spatial resolution is defined by the number of pixels per side of the unit cell. The filter radius is defined with respect to the pixel size: a radius of 0.5 corresponds to the inscribe circle of a square pixel. Finally, the approximate interface, required for the 2-layer filter, is defined at grid points $x_\alpha$ by its normal vector linking the inclusion center to $x_\alpha$.

3 – Improving spatial convergence

The spatial convergence is studied on the average axial stress evaluated on a set of simulations performed with spatial resolutions between 6 and 128, with various filters, filter radii and contrasts (figures 1-A and C). These curves are used to evaluate $N_{1\%}$, the minimum spatial resolution required to obtain a precision less than 1% on the average axial stress, which is plotted as a function of the filter radius (figures 1-B and D).

The first significant effect concerns the erratic spatial convergence observed without any filter (figures 1-A and C). When considering the 2-layer or the 'well chosen' (see just below) Voigt and Reuss filters, the convergence curves exhibit a much smoother evolution. This point is of a practical interest in order to optimize the spatial resolution with only a few simulations, avoiding the time-consuming building of figure 1 (one simulation per resolution!).

Secondly, it should be mentioned that the use of the Voigt and Reuss filters strongly depends on the elastic contrast. Actually, significant improvements are obtained when choosing the filter in agreement with the matrix behavior: a compliant homogenization rule (Reuss) with a compliant matrix and a stiff one (Voigt) with a stiff matrix. On the contrary, inverting this choice significantly deteriorates the results. The 2-layer filter proposed in this note is a bit less efficient than the Voigt and Reuss filters, when chosen carefully, but proves to be efficient in both situations and is then expected to be of a great interest for more complex microstructures (polycrystals, bi-percolated microstructures etc...).

Finally, figure 1-B and D reveal that quasi-identical optimum filter radii (between 0.5 and 0.55) can be found for the two efficient filters and the two elastic contrasts. For this optimum value, the minimum required resolution $N_{1\%}$ can be divided by a factor between 2.6 and 7.2 . Hence, the problem size is divided by a factor between 6.7 and 51.8, here in 2D (and between 17.5 and 373 if extrapolated in 3D). As a result, filtering material properties sharply reduces the problem size and consequently the computation time without any modification of the iterative algorithm or the discrete Green operator.

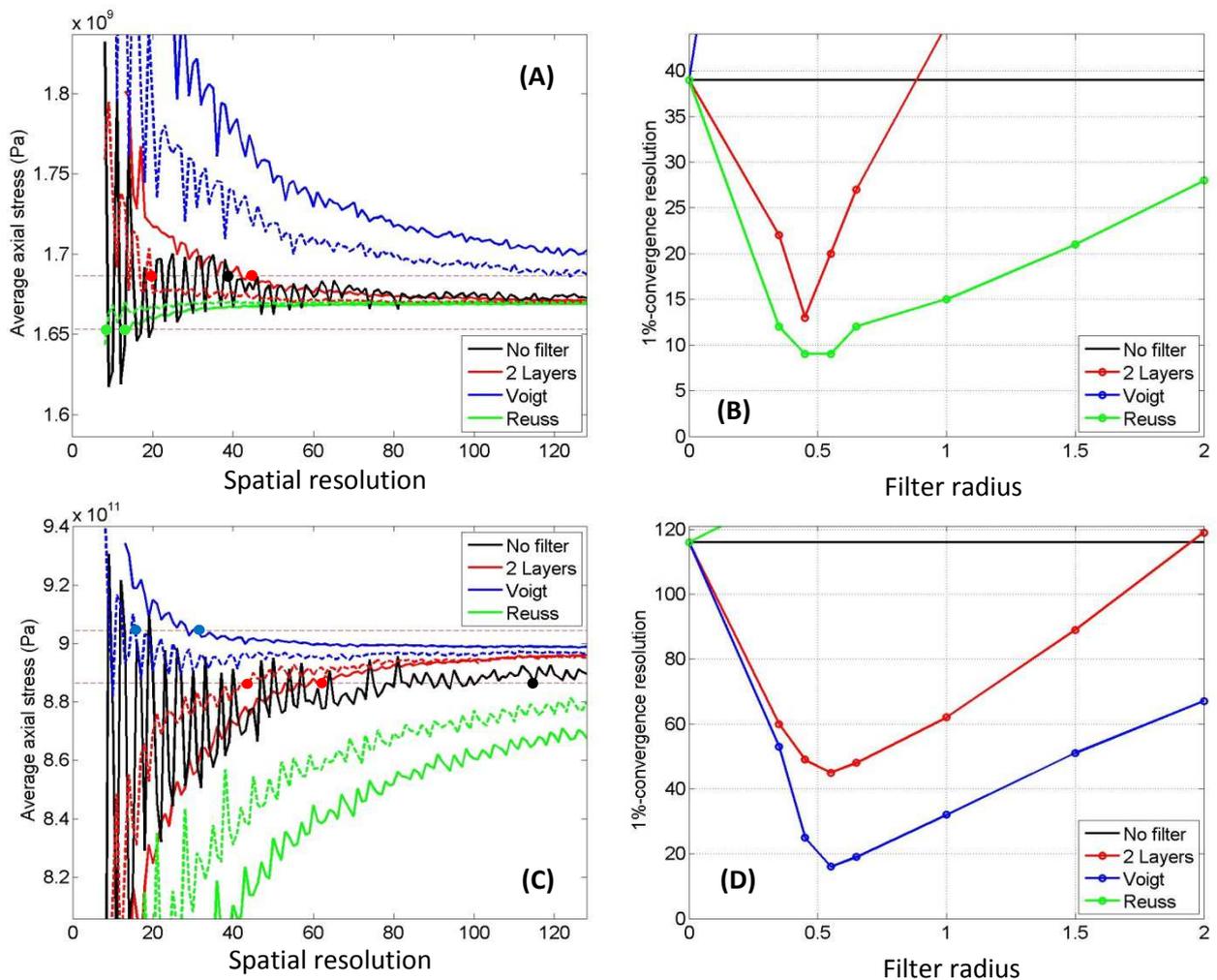

Figure 1: A and C plot the evolution of the average axial stress as a function of the spatial resolution* with elastic contrasts* of $10^3$ and $10^{-3}$ respectively (inverted behaviors). Solid and dashed lines correspond respectively to filter radii* of 0.55 and 1.0 and the horizontal red dashed lines represent the +/-1% converged average axial stresses. For each contrast, B and D plot the evolution of the minimum required resolution $N_{1\%}$ (materialized by colored points on A and C for filter radii of 0.55 and 1.0) as a function of the filter radius. (*: see definition at the end of section 2).

4 – Reducing spurious oscillations

The observation of the stress maps on this numerical experiment (figure 2) evidences that spurious oscillations can be reduced by filtering material properties.
Actually, one of the well-known drawbacks of these spectral methods is the development of spurious oscillations on the stress and strain fields, especially visible with highly contrasted material properties (figure A). Surprisingly, the optimal filter radius, 0.55 for a contrast of $10^{-3}$, leads to a much better estimate of the macroscopic behavior (figure 1-D) whereas no significant effect are observed on the stress fields (figure 2-C and D). In a similar way, the difference between the 2-layer and the Voigt filters observed macroscopically (figure 1-D) is almost invisible on the stress maps. To obtain a visible reduction of spurious oscillations (figure 2-E and F), the filter radius is increased from 0.55 to 1.0 (a circle inscribed in a 2x2 pixels square). This arbitrary value can be regarded as a compromise between the reduction of spatial oscillations and the reduction of minimum required resolution $N_{1\%}$ (on figure 1-D, $N_{1\%}$ is still from 2 to 4 times lower than without filter).

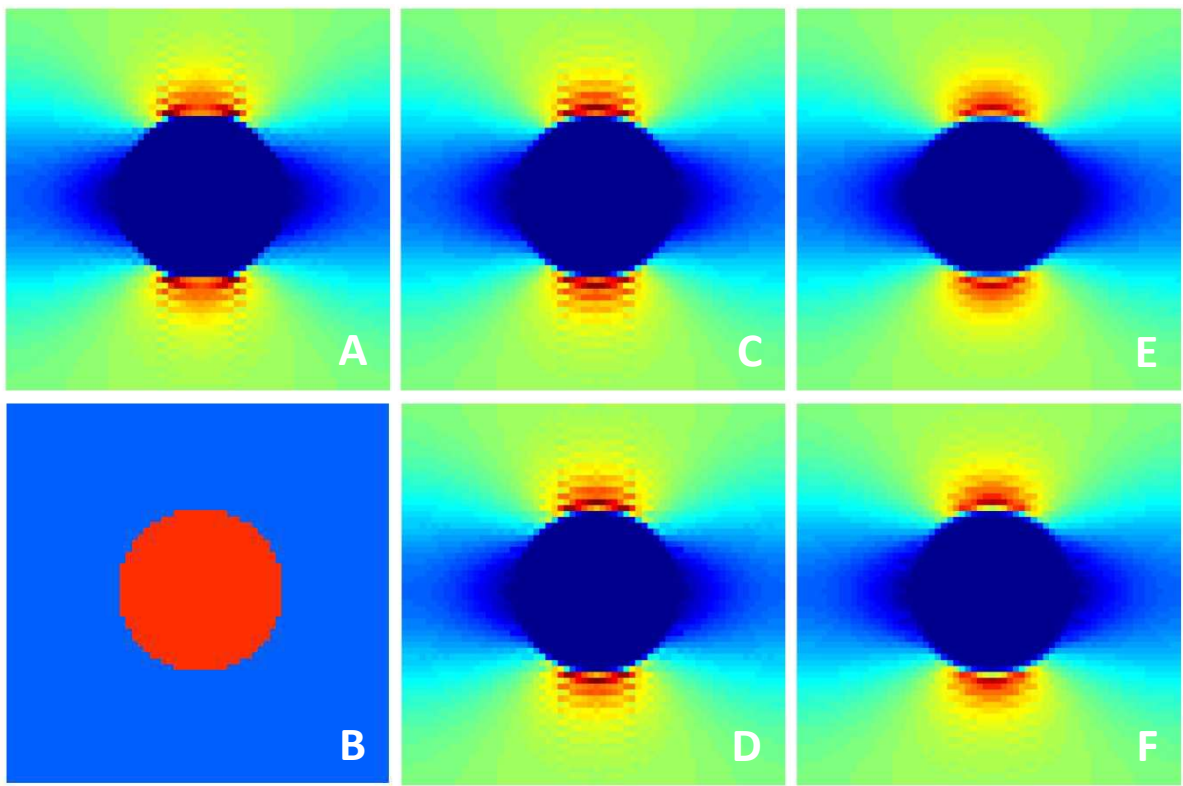

Figure 2 : Axial stress maps (horizontal uniaxial loading axis) for the unit cell discretized with a 65 spatial resolution (B) and a $10^{-3}$ elastic contrast*: without any filter (A), with Voigt (C - E) and 2-layer (D-F) homogenization rules associated to filter radii* of 0.55 (C-D) and 1.0 (E-F). (*: see definition at the end of section 2).

5- Sensitivity to elastic contrast and reference material

The other two drawbacks of the fixed-point algorithms are their high sensitivity to the reference material and to the material contrast [7]. It must be mentioned that Krylov subspace algorithms (such as the Conjugate Gradient) almost vanishes the first one and sharply reduces the second [12]. Figure 3 demonstrates, here with the classical fixed-point algorithm (section 2), that filtering material properties does not have positive nor negative effect on these two drawbacks.

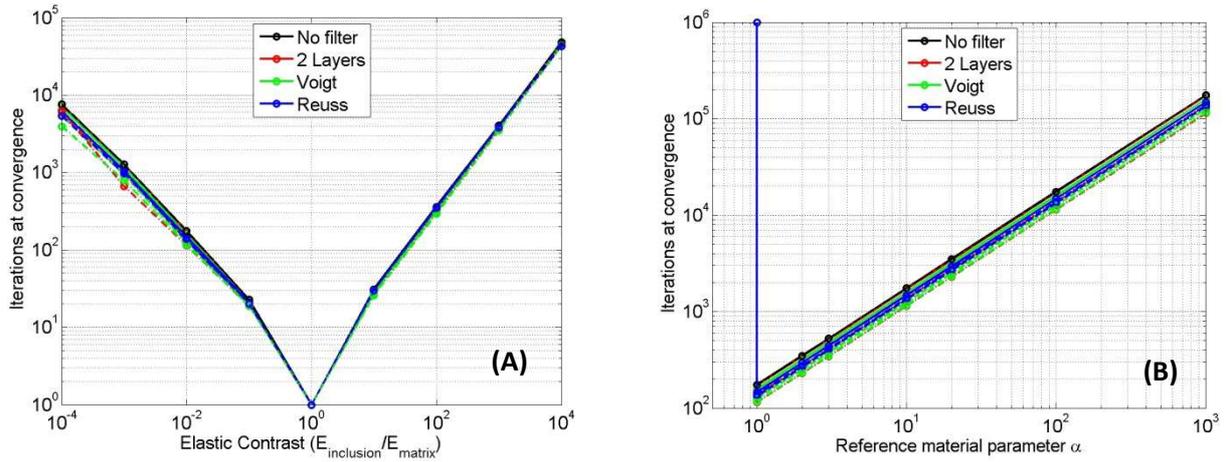

Figure 3: Number of iterations at convergence (spatial resolution=65) as a function of the elastic contrast (A) and of the material reference (B). The parameter α multiplies the optimal choice given in [9] for isotropic elastic phases. Below this optimal value, the algorithm does not converge. Solid, dashed and dashed-dot lines, associated to 0.55, 1.0 and 1.5 filter radius, are almost superimposed.

6- Conclusion and discussion

This short note demonstrates on a simple numerical experiment how filtering material properties is able to efficiently improve the spectral methods applied to heterogeneous materials, essentially by reducing the required spatial resolution and/or reducing spurious oscillations. This approach only modifies the input parameters of these methods and can be used in addition to improved algorithms or modified discrete Green operators. In contrast with Voigt and Reuss filters, the 2-layer filter, proposed in this note, is efficient in both cases of a stiff or compliant inclusion within a matrix.

The choice of the filter radius should be at least the optimal value found on figure 1 (B and D), almost 0.5 (approximately the size of a pixel, or voxel in 3D), which significantly decreases the minimum required spatial resolution $N_{1\%}$ but without any significant effect on the local fields. Increasing the filter radius reduces spurious oscillations but increases the minimum required resolution $N_{1\%}$. Hence, the choice will result from a compromise between the computation time (closely related to the resolution) and the quality of the local fields.

These conclusions are discussed in the view of extrapolated material properties obtained from truncated Fourier series with the Fourier coefficients evaluated from FFT applied to the fields $\widetilde{c}(x_\alpha)$ (Figure 4). Without any filter, important oscillations are observed and a part of this oscillations seems to correspond to spurious oscillations observed on figure 2-A. For a radius of 0.55, oscillations are reduced but still visible, and for a radius of 1.0, oscillations are almost invisible in line with the reduced spurious oscillations on figure 2-E and F. On the other hand, increasing the filter radius increases the 'interphase' thickness and decreases the agreement with the real microstructure leading to increase the required spatial resolution (Figure 1). Finally, minimizing the gap between the exact microstructure and extrapolated ones lead to an optimal value of 0.55, consistent with the optimal value deduced from numerical simulations. To conclude this discussion, the Fourier extrapolation of material properties could explain our results in terms of spurious oscillations and evolution of the minimum required resolution $N_{1\%}$.

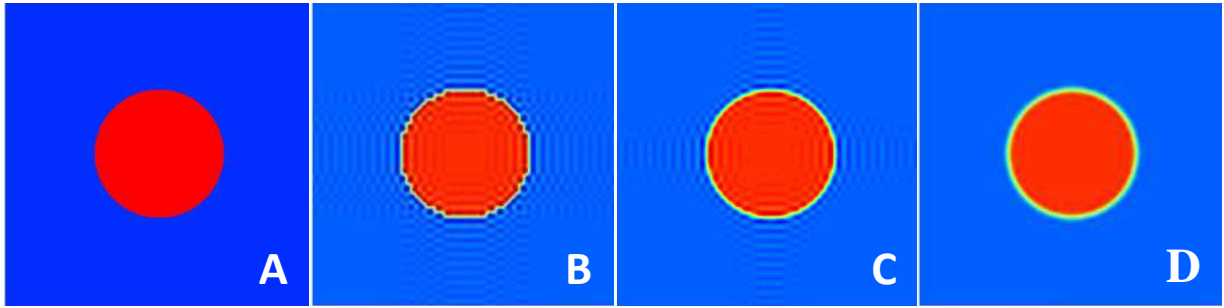

Figure 4: 'Exact' material property (A). Extrapolation with truncated Fourier series associated to a 65x65 grid of the material properties without any filter (B) and with filter radii of 0.55 (C) and 1.0 (D).